\documentclass[preprint,nofootinbib]{revtex4}

\usepackage{amsmath}
\usepackage{graphicx}

\newcommand{\beq}{\begin{equation}}
\newcommand{\eeq}{\end{equation}}
\newcommand{\beqa}{\begin{eqnarray}}
\newcommand{\eeqa}{\end{eqnarray}}
\newcommand{\no}{\nonumber}
\def\OMIT#1{{}}
\newcommand{\lsim}{\mathrel{\rlap{\lower4pt\hbox{\hskip1pt$\sim$}}
    \raise1pt\hbox{$<$}}}         
\newcommand{\gsim}{\mathrel{\rlap{\lower4pt\hbox{\hskip1pt$\sim$}}
    \raise1pt\hbox{$>$}}}         

\begin{document}


\vspace*{.0cm}

\title{Lessons from BaBar and Belle measurements\\ of
  $D^0-\overline{D^0}$ mixing parameters}

\author{Yosef Nir\footnote{The Amos de-Shalit chair of theoretical physics}}\email{yosef.nir@weizmann.ac.il}
\affiliation{Department of Particle Physics,
  Weizmann Institute of Science, Rehovot 76100, Israel\vspace*{3mm}}


\begin{abstract}\vspace*{3mm}
  The BaBar and Belle experiments have recently presented evidence for
  $D^0-\overline{D^0}$ mixing. We explain the following points: (i)
  The measurements imply width difference $y\sim0.01$. In the limit of
  small CP violation, the CP-odd state is longer-lived; (ii)
  $y\sim0.01$ is consistent with the Standard Model. It suggests that
  SU(3) breaking from phase space effects is likely to play a major
  role; (iii) There is no evidence for either large mass splitting or
  CP violation. Consequently, there is no hint for new physics; (iv)
  The stronger bounds on the mass splitting and on CP violation imply
  that, if squarks are observed at the LHC, it is unlikely that they
  will be non-degenerate.
\end{abstract}

\maketitle

\section {Introduction}
\label{intro}
Neutral meson mixing has been observed in all down-type neutral meson
systems ($K$, $B$ and $B_s$) providing a sensitive probe of the
flavor structure of the Standard Model and its extensions. In
contrast, mixing in the neutral $D$-meson system (the {\it only}
up-type neutral meson)
has not been observed until very recently. This situation is now
changed. Measurements of the doubly Cabibbo suppressed (DCS) $D^0\to
K^+\pi^-$ decay by the BaBar experiment \cite{Aubert:2007wf}, and of
the singly Cabibbo suppressed (SCS) $D^0\to K^+K^-,\pi^+\pi^-$ decays
by the Belle experiment \cite{Abe:2007dt}, have given evidence of width
difference between the two neutral $D$-meson mass eigenstates:
\beqa
y^\prime\cos\phi&=&(0.97\pm0.44\pm0.31)\times10^{-2},\\
y_{\rm CP}&=&(1.31\pm0.32\pm0.25)\times10^{-2}.
\eeqa
In this note we explain the significance of these results.

In section \ref{sec:for} we present the formalism of DCS and SCS
neutral $D$ decays, and the simplifications that follow from
neglecting direct CP violation. In section \ref{sec:inter} we
interpret the new results qualitatively and quantitatively and explain
their implications for width difference, mass difference, CP violation
and strong phases in the neutral $D$-meson system. In section
\ref{sec:models} we examine the implications of the new results for
the Standard Model, and to models of new physics, particularly
supersymmetry with alignment. We summarize our conclusions in section
\ref{sec:con}. We collect new and previous relevant experimental
results and derive world averages in Appendix \ref{app:exp}.

\section {Formalism}
\label{sec:for}
In this section, we present the formalism that describes the neutral
$D$ decay and mixing, following the analysis of
Ref. \cite{Bergmann:2000id}. The two neutral $D$-meson mass
eigenstates, $|D_1\rangle$ of mass $m_1$ and width $\Gamma_1$ and
$|D_2\rangle$ of mass $m_2$ and width $\Gamma_2$ are linear
combinations of the interaction eigenstates $D^0$ (with quark content
$c\overline{u}$) and $\overline{D^0}$ (with quark content
$\overline{c}u$):
\beqa\label{masses}
|D_1\rangle&=&p|D^0\rangle+q|\overline{D^0}\rangle,\no\\
|D_2\rangle&=&p|D^0\rangle-q|\overline{D^0}\rangle.
\eeqa
The average and the difference in mass and width are given by 
\beqa\label{sumdif}
m\equiv\frac{m_1+m_2}{2},&\ \ \
&\Gamma\equiv\frac{\Gamma_1+\Gamma_2}{2},\no\\
x\equiv\frac{m_2-m_1}{\Gamma},&\ \
&y\equiv\frac{\Gamma_2-\Gamma_1}{2\Gamma}.
\eeqa
The decay amplitudes into a final state $f$ are defined as follows: 
\beqa\label{decays}
A_f&=&\langle f|{\cal H}|D^0\rangle,\no\\
\overline{A}_f&=&\langle f|{\cal H}|\overline{D^0}\rangle.
\eeqa
We define $\lambda_f$:
\beq\label{deflam}
\lambda_f=\frac{q}{p}\frac{\bar A_f}{A_f}.
\eeq

We now write the approximate expressions for the time-dependent DCS
and SCS decay rates that are valid for time $t\lsim1/\Gamma$. We take
into account the experimental information that $x$, $y$ and
$\tan\theta_c$ (where $\theta_c$ is the Cabibbo angle) are small, and
expand each of the rates only to the order that is relevant to the
BaBar and Belle measurements: 
\beqa\label{dcsgen}
\Gamma[D^0(t)&\to&K^+\pi^-]=e^{-\Gamma
  t}|\overline{A}_{K^+\pi^-}|^2|q/p|^2\no\\
&\times&\left\{|\lambda_{K^+\pi^-}^{-1}|^2
  +[{\cal R}e(\lambda_{K^+\pi^-}^{-1})y
  +{\cal I}m(\lambda_{K^+\pi^-}^{-1})x]\Gamma t
  +\frac14(y^2+x^2)(\Gamma t)^2\right\},\no\\
\Gamma[\overline{D^0}(t)&\to&K^-\pi^+]=e^{-\Gamma
  t}|{A}_{K^-\pi^+}|^2|p/q|^2\\
&\times&\left\{|\lambda_{K^-\pi^+}|^2
  +[{\cal R}e(\lambda_{K^-\pi^+})y
  +{\cal I}m(\lambda_{K^-\pi^+})x]\Gamma t
  +\frac14(y^2+x^2)(\Gamma t)^2\right\},\no
\eeqa
\beqa\label{scsgen}
\Gamma[D^0(t) \to K^+K^-]&=&e^{-\Gamma t}|{A}_{K^+K^-}|^2
\left\{1+[{\cal R}e(\lambda_{K^+K^-})y
  -{\cal I}m(\lambda_{K^+K^-})x]\Gamma t\right\},\no\\
\Gamma[\overline{D^0}(t)\to K^+K^-]&=&e^{-\Gamma
  t}|\overline{A}_{K^+K^-}|^2 
\left\{1+[{\cal R}e(\lambda_{K^+K^-}^{-1})y
  -{\cal I}m(\lambda_{K^+K^-}^{-1})x]\Gamma t\right\}.
\eeqa

Within the Standard Model, the physics of $D^0-\overline{D^0}$ mixing
and of the tree level decays is dominated by the first two generations
and, consequently, CP violation can be safely neglected (for reviews
of charm physics, see \cite{Bianco:2003vb,Burdman:2003rs}). Indeed, CP
violation in these processes would constitute a signal for new physics
\cite{Bigi:1986dp,Blaylock:1995ay,Bergmann:2000id}. In all
`reasonable' extensions of the Standard Model, both the DCS
\cite{Bergmann:1999pm} and the SCS \cite{Grossman:2006jg} decays are
still dominated by the Standard Model CP conserving contributions. On
the other hand, there could be new short distance, possibly CP
violating contributions to the mixing amplitude $M_{12}$. Allowing for
only such CP violating effects of new physics, the picture of CP
violation is simplified since there is no direct CP
violation.\footnote{In some supersymmetric models, SCS decays may
  exhibit comparable direct and indirect CP violations
  \cite{Grossman:2006jg}.} The effects of indirect CP violation can be
parametrized in the following way \cite{Nir:1999mg}:
\beqa
\lambda^{-1}_{K^+\pi^-}&=&r_d|p/q|e^{-i(\delta+\phi)},\no\\
\lambda_{K^-\pi^+}&=&r_d |q/p|e^{-i(\delta-\phi)},\no\\
\lambda_{K^+K^-}&=&-|q/p|e^{i\phi},
\eeqa
where $r_d$ is a real and positive dimensionless parameter,
$\delta$ is a strong (CP conserving) phase, and $\phi$ is a weak (CP
violating) phase. The appearance of a single weak phase common to all
final states is related to the absence of direct CP violation, while
the absence of a strong phase in $\lambda_{K^+K^-}$ is related to the
fact that the final state is a CP eigenstate. CP violation in mixing
is related to
\beq
A_m\equiv\frac{|q/p|^2-1}{|q/p|^2+1}\neq0.
\eeq
CP violation in the interference of decays with and without mixing is
related to $\sin\phi\neq0$. In the limit of CP conservation, where the
mass eigenstates are also CP eigenstates, choosing $\phi=0$ is
equivalent to defining $|D_1\rangle=|D_-\rangle$ and
$|D_2\rangle=|D_+\rangle$, with $D_-(D_+)$ being the CP-odd (CP-even)
state, that is, the state that does not (does) decay into $K^+K^-$.
(Alternatively, $\phi=\pi$ is also a legitimate choice in the CP
conserving case; it simply identifies $|D_1\rangle=|D_+\rangle$ and
$|D_2\rangle=|D_-\rangle$. The physical observable $y\cos\phi$ remains
unchanged under these alternative conventions.)

For the analysis of the DCS decays, it is convenient to further define
\beqa\label{defxyp}
x^\prime&\equiv& x\cos\delta+y\sin\delta,\no\\
y^\prime&\equiv& y\cos\delta-x\sin\delta.
\eeqa
In the absence of direct CP violation, the expressions for the DCS decay rates
(\ref{dcsgen}) and for the SCS decay rates (\ref{scsgen}) simplify:
\beqa\label{dcsnod}
\Gamma[D^0(t)&\to&K^+\pi^-]=e^{-\Gamma
  t}|{A}_{K^-\pi^+}|^2\no\\
&\times&\left[r_d^2+r_d|q/p|(y^\prime\cos\phi-x^\prime\sin\phi)\Gamma t
  +\frac14{|q/p|^2}(y^2+x^2)(\Gamma t)^2\right],\no\\
\Gamma[\overline{D^0}(t)&\to&K^-\pi^+]=e^{-\Gamma
  t}|{A}_{K^-\pi^+}|^2\\
&\times&\left[r_d^2+r_d|p/q|(y^\prime\cos\phi+x^\prime\sin\phi)\Gamma t
  +\frac14{|p/q|^2}(y^2+x^2)(\Gamma t)^2\right],\no
\eeqa
\beqa\label{scsnod}
\Gamma[D^0(t) \to K^+K^-]&=&e^{-\Gamma t}|{A}_{K^+K^-}|^2
\left[1-|q/p|(y\cos\phi-x\sin\phi)\Gamma t\right],\no\\
\Gamma[\overline{D^0}(t)\to K^+K^-]&=&e^{-\Gamma t}|{A}_{K^+K^-}|^2 
\left[1-|p/q|(y\cos\phi+x\sin\phi)\Gamma t\right].
\eeqa

Ref. \cite{Aubert:2007wf} uses parameters $y^\prime_\pm$ and
$x^{\prime2}_\pm$ that correspond to the following combinations of
parameters: 
\beqa
y^\prime_+=|q/p|(y^\prime\cos\phi-x^\prime\sin\phi),&\ \ \
&x^\prime_+=|q/p|(x^\prime\cos\phi+y^\prime\sin\phi),\no\\
y^\prime_-=|p/q|(y^\prime\cos\phi+x^\prime\sin\phi),&\ \ \
&x^\prime_-=|p/q|(x^\prime\cos\phi-y^\prime\sin\phi).
\eeqa
In the limit of CP conservation,
\beqa\label{yprze}
y^\prime_+&=&y^\prime_-\equiv y^\prime_0=
\left(\frac{\Gamma_+-\Gamma_-}{2\Gamma}\right)\cos\delta
- \left(\frac{m_+-m_-}{\Gamma}\right)\sin\delta,\no\\
x^\prime_+&=&x^\prime_-\equiv x^\prime_0=
\left(\frac{\Gamma_+-\Gamma_-}{2\Gamma}\right)\sin\delta
+ \left(\frac{m_+-m_-}{\Gamma}\right)\cos\delta,
\eeqa
where sub-indices $+(-)$ in $\Gamma_\pm$ and $m_\pm$ denote the
CP-even (-odd) mass eigenstate. 

Ref. \cite{Abe:2007dt} uses parameters $y_{\rm CP}$ and $A_\Gamma$ that
correspond to the following combinations of parameters:\footnote{In
  the notations of the PDG \cite{Yao:2006px}, $y_{\rm CP}\equiv Y$ and
  $A_\Gamma\equiv-\Delta Y$.}
\beqa\label{ycp}
y_{\rm CP}&=&\frac12(|q/p|+|p/q|)y\cos\phi
-\frac12(|q/p|-|p/q|)x\sin\phi,\\
\label{agamma}
A_\Gamma&=&\frac12(|q/p|-|p/q|)y\cos\phi
-\frac12(|q/p|+|p/q|)x\sin\phi.
\eeqa
In the limit of CP conservation,
\beqa
y_{\rm CP}&=&\frac{\Gamma_+-\Gamma_-}{2\Gamma},\no\\
A_\Gamma&=&0.
\eeqa

\section{Interpreting the data (model independently)}
\label{sec:inter}
Ref. \cite{Abe:2007dt} gives the following results related to the SCS
decays: 
\beqa\label{belleycp}
y_{\rm CP}&=&(1.31\pm0.32\pm0.25)\times10^{-2},\\
\label{belleag}
A_\Gamma&=&(0.01\pm0.30\pm0.15)\times10^{-2}.
\eeqa

Two straightforward statements follow from Eqs. (\ref{belleycp})
and (\ref{belleag}):
\begin{itemize}
\item There is evidence for $D^0-\overline{D^0}$ mixing;
\item There is no evidence for CP violation in $D^0-\overline{D^0}$
  mixing.
\end{itemize}

We would like to be more quantitative on the issue of CP violation. If
we make the plausible assumption that $y_{\rm CP}$ is dominated by
the $y$-term (note that the $x$-term has two CP violating factors,
while the $y$-term has none), then the ratio $A_\Gamma/y_{\rm CP}$ is
very informative: 
\beq\label{ratay}
\frac{A_\Gamma}{y_{\rm CP}}\approx A_m-\frac xy \tan\phi.
\eeq
It will be helpful if experiments quote their results directly
for this ratio. We assume that the systematic errors in Eqs.
(\ref{belleycp}), (\ref{belleag}) cancel in this ratio. Then we obtain
the following constraint on CP violation:
\beq\label{besmallcp}
A_m-\frac xy \tan\phi\sim0.0\pm0.3.
\eeq
There could be cancellations between the two terms in
Eq. (\ref{besmallcp}). Furthermore, at the $2-3\sigma$ level, CP
violation could still be large. Yet, barring fine-tuned cancellations,
the results are suggestive (at the $1-2\sigma$ level) that
$|q/p|\approx1$ and either $|\sin\phi|<1$ or $|x|<|y|$ (or
both). If the correct interpretation of (\ref{belleag}) is indeed that
CP violation is small, then (\ref{belleycp}) reads 
\beq
\frac{\Gamma_+-\Gamma_-}{2\Gamma}\approx(+1.3\pm0.4)\times10^{-2}. 
\eeq

To summarize, the Belle results are suggestive of the following
statements regarding $D^0-\overline{D^0}$ mixing:
\begin{itemize}
\item The width difference is of order one percent;
  \item The CP-odd state is longer-lived;
\item CP violation in mixing is small;
\item Either the mass difference is smaller than the width difference,
  or CP violation in the interference of decays with and without
  mixing is small, or both.
\end{itemize}

Ref. \cite{Aubert:2007wf} gives the following results related to the DCS
decays:\footnote{Ref. \cite{Aubert:2007wf} allows for direct CP
  violation in their fit. The results are, however, consistent with
  vanishing direct CP violation. We therefore use our formalism which
  neglects direct CP violation.} 
\beqa\label{babarcpv}
y_+^\prime&=&(0.98\pm0.64\pm0.45)\times10^{-2},\no\\
y_-^\prime&=&(0.96\pm0.61\pm0.43)\times10^{-2},\no\\
x_+^{\prime2}&=&(-2.4\pm4.3\pm3.0)\times10^{-4},\no\\
x_-^{\prime2}&=&(-2.0\pm4.1\pm2.9)\times10^{-4}.
\eeqa
The fact that the results are consistent with $y^\prime_+=y^\prime_-$
means that also here there is no evidence for CP violation. Making the
plausible assumption that $y^\prime_\pm$ are
dominated by the $y^\prime\cos\phi$ terms, then the ratio
$(y^\prime_+-y^\prime_-)/(y^\prime_++y^\prime_-)$ can be simply
interpreted:   
\beq\label{ratypm}
\frac{y^\prime_+-y^\prime_-}{y^\prime_++y^\prime_-}\approx
A_m-\frac{x^\prime}{y^\prime}\tan\phi.
\eeq
Again, it will be helpful if experiments quote their results directly
for this ratio. We assume that the systematic errors in Eqs.
(\ref{babarcpv}) cancel in this ratio. We further
assume that there are no fine-tuned cancellations between the two
terms in Eq. (\ref{ratypm}). Then we obtain upper bounds on CP violation
that are somewhat weaker than Eq. (\ref{besmallcp}), namely 
$A_m-({x^\prime}/{y^\prime}) \tan\phi\sim0.0\pm0.6$.

Neglecting CP violation, Ref. \cite{Aubert:2007wf} obtains the following fit:
\beqa\label{babarcpc}
y^\prime_0&=&(0.97\pm0.54)\times10^{-2},\\
x^{\prime2}_0&=&(-2.2\pm3.7)\times10^{-4}.
\eeqa
where $y^\prime_0=y$ for $\phi=0$ [see Eq. (\ref{yprze})]. Taking into
account the strong correlation between these two observables, BaBar
finds that the possibility of no mixing is disfavored at the 3.9
standard deviations. It is interesting to note, however, that Belle
finds \cite{Zhang:2006dp}, for the CP conserving case,
$y^\prime_0=(0.06\pm0.40)\times10^{-2}$. The average of the two
results for $y^\prime_0$ (see Appendix \ref{app:exp}) does not show
evidence for mixing, but one must remember that this simple averaging
does not keep the correlation information.

The results on SCS and DCS decays cannot be combined in a
straightforward way, because of the presence of strong phases in the
$D\to K\pi$ decays. Only if one assumes that the strong phase is
small, one can interpret the BaBar result in terms of $y$. Indeed,
$\delta=0$ in the flavor SU(3) limit, but it is not clear whether the
relevant SU(3) breaking effects are small
\cite{Chau:1993ec,Browder:1995ay} or large \cite{Falk:1999ts}. In any
case, as long as $\delta<\pi/2$, the BaBar result is also suggestive
that the CP-even state has a shorter lifetime.

Actually, we can use the experimental results to get an idea about the
strong phase $\delta$.  Following \cite{Bergmann:2000id}, we
approximate $|q/p|^2=1$ and $|\sin\phi|=0$ and combine (\ref{defxyp})
and (\ref{ycp}) to get
\beq\label{findcd}
\frac{y^\prime\cos\phi}{y_{\rm CP}}=\cos\delta-\frac{x}{y}\sin\delta.
\eeq
If we consider only the recent Belle result for $y_{\rm CP}$ and BaBar
result for $y^\prime\cos\phi$, we obtain
\beq
+0.74\pm0.47=\cos\delta-(x/y)\sin\delta.
\eeq
For $|x|\ll|y|$ we have $|\delta|\lsim5\pi/12$. For $x\sim-y$, the
preferred ranges are around $\delta\sim0$ or $\delta\sim\pi/2$ while
for $x\sim +y$, the preferred ranges are around $\delta\sim0$ or
$\delta\sim3\pi/2$. In all cases, the results are consistent with
$\delta=0$. On the other hand, if we use the world averages
(\ref{ycpwa}) for $y_{\rm CP}$ and (\ref{ypwa}) for
$y^\prime\cos\phi$, then the range for the left hand side is lower,
roughly $0.2\pm0.2$, and $\delta=0$ (or, more generally,
$\cos\delta\gsim0.9$) is disfavored. We conclude that more data is
needed to clarify the situation regarding the strong phase.

\section{Implications for models}
\label{sec:models}
\subsection{The Standard Model}
Because of the GIM mechanism, the mixing amplitude is proportional to
differences of terms suppressed by $m^2_{d,s,b}/m^2_W$, and so
$D^0-\overline{D^0}$ is very slow in the Standard Model (for a survey
of predictions, see \cite{Nelson:1999fg,Petrov:2006nc}).  The
contribution of the $b$ quark is further suppressed by the small CKM
elements $|V_{ub}V_{cb}^*|^2/|V_{us}V_{cs}^*|^2= {\cal O}(10^{-6})$,
and can be neglected. Thus, the $D$ system essentially involves only
the first two generations, and therefore CP violation is absent both
in the mixing amplitude and in the dominant tree-level decay
amplitudes. Once the contribution of the $b$ quark is neglected, the
mixing vanishes in the flavor SU(3) limit and, if SU(3) breaking can
be treated analytically, it only arises at second order in SU(3)
breaking \cite{Falk:2001hx,Falk:2004wg}:
\beq\label{sutb}
x,y\sim\sin^2\theta_c\times[SU(3)\ {\rm breaking}]^2.
\eeq
Precise calculations of $x$ and $y$ in the Standard Model are not
possible at present, because the charm mass is neither heavy enough to
justify inclusive calculations
\cite{Georgi:1992as,Ohl:1992sr,Bigi:2000wn}, nor is it light enough 
to allow a few exclusive channels to give a reliable estimate
\cite{Donoghue:1985hh,Buccella:1996uy,Golowich:1998pz,Wolfenstein:1985ft,Colangelo:1990hj,Kaeding:1995zx}.  
Most studies (particularly the `inclusive' ones) find
$x,y\lsim10^{-3}$. Ref. \cite{Bigi:2000wn} raises the possibility that
$x,y$ will be measured at the $10^{-2}$ level, interpreting such a
result as breakdown of the OPE.

According to Eq. (\ref{sutb}), computing $x$ and $y$ in the Standard
Model requires a calculation of SU(3) violation in decay rates. There
are many sources of SU(3) violation, most of them involving
nonperturbative physics in an essential way. In
Ref. \cite{Falk:2001hx}, SU(3) breaking arising from phase space
differences was studied; computing them in two-, three-, and four-body
$D$ decays, it was found that $y$ could naturally be at the level of
one percent. The result can be traced back to the fact that the SU(3)
cancellation between the contributions of members of the same
multiplet can be badly broken when decays to the heaviest members of
a multiplet have a small or vanishing phase space. This effect is
manifestly not included in the OPE-based calculations of
$D^0-\overline{D^0}$ mixing, which cannot address threshold
effects. The experimental results, implying $y={\cal O}(0.01)$,
suggest that the phase space effect analyzed in
Ref. \cite{Falk:2001hx} is, very likely, a significant if not the
dominant source of the width splitting. In particular, there is no
significant cancellation against other sources of SU(3) breaking.

If the dominant SU(3) breaking mechanism is indeed the one studied in
Ref. \cite{Falk:2001hx}, should we expect $x$ to be comparably large?
The task of answering this question was taken in Ref.
\cite{Falk:2004wg}. The Standard Model prediction for $x/y$ due to
SU(3) breaking from final state phase space differences was studied. A
dispersion relation relating $\Delta m$ to $\Delta\Gamma$ using Heavy
Quark Effective Theory (HQET) was derived. The calculation is less
model independent than the one of $y$ \cite{Falk:2001hx}, and should
be trusted only at the order of magnitude level. The final conclusion
was that, if $y$ is dominated by the four body decays considered in
\cite{Falk:2001hx}, we should expect $|x|$ between $10^{-3}$ and
$10^{-2}$, and that $x$ and $y$ are of opposite signs.

We conclude that the evidence for $y\sim0.01$, the upper bound on
$x\lsim0.02$, and the absence of signals of CP violation are all very
consistent with the Standard Model. The value of $y$ implies large
SU(3) breaking effects, of just the right size to be accounted for by
phase space effects identified in \cite{Falk:2001hx}.

\subsection{Beyond the Standard Model}
New physics modifies the $\Delta C=2$ part in the $D^0-\overline{D^0}$
mixing amplitude, $M_{12}^D$. It could give mixing that is close to
the experimental bound. This situation is unavoidable in
supersymmetric models where the only flavor suppression mechanism is
alignment \cite{Nir:1993mx,Leurer:1993gy,Nir:2002ah}. Hence, it is
important to find the precise limit on $M_{12}^D$.

In terms of measurable quantities,
$|M_{12}^D|$ is given by \cite{Branco:1999fs,Raz:2002ms}
\beq
|M_{12}^D|^2=\left(\frac{x\Gamma}{2}\right)^2\
\frac{1+A_m^2(y/x)^2}{1-A_m^2}.
\eeq
The strongest bound would apply in the CP conserving case, $|q/p|=1$,
in which case $|M_{12}^D|=|x|\Gamma/2$. Using the new Belle result of
Eq. (\ref{belledalitz}) to obtain an upper bound $|x|\lsim0.015$ (95\%
C.L.), we get
\beq
|M_{12}^D|\lsim1.2\times10^{-11}\ MeV\ \ ({\rm CP\ conservation}),
\eeq
a factor of two stronger than \cite{Raz:2002ms}. The bound becomes,
however, weaker in the presence of CP violation in mixing. If we take 
$A_m^2\lsim0.3$ [see Eq. (\ref{besmallcp})], then the bound is
relaxed by a factor $\sim2$:
\beq
|M_{12}^D|\lsim2.2\times10^{-11}\ MeV\ \ ({\rm CP\ violation}),
\eeq
a factor of three stronger than \cite{Raz:2002ms}. A numerical fit of
the five relevant parameters ($y,x,\delta,\phi,|q/p|$) to the six
observables ($y_{\rm CP},A_\Gamma,y^\prime_\pm,x^\prime_\pm$) will
give more accurate results. It will be useful, however, for the
purpose of such fit, if experiments quote the errors directly on the
CP violating ratios (\ref{ratay}) and (\ref{ratypm}).

When the bound on $|M_{12}^D|$ is strengthened by a factor $\sim3$,
the bound on the relevant flavor changing supersymmetric parameter
$(\delta^u_{LL})_{12}$ is strengthened by a factor $\sim\sqrt{3}$.
Barring accidental cancellations between the Standard Model and the
supersymmetric contribution, or between various terms in the
supersymmetric contribution (these appear for a certain ratio between
the squark and the gluino masses), or RGE-induced approximate
degeneracy \cite{Nir:2002ah}, this stronger bound can be translated
into a lower bound on the scale of gluino and up-squark masses of
order 2 TeV, which is uncomfortably high.

Alignment models have provided the only natural example of models with
a squark spectrum that is potentially both light and non-degenerate.
Consequently, the lessons from the new constraints can be stated as
follows: 
\begin{itemize}
  \item If squark masses are within the reach of the LHC, it is very
    unlikely that they will show no degeneracy.
    \item If such a situation is nevertheless realized in Nature, it
      requires a specific relation between the up-squark and gluino
      mass \cite{Nir:2002ah} or accidental strong cancellations
      between the Standard Model and the supersymmetric contributions.
      \end{itemize}

\section{Conclusions}
\label{sec:con}      
\begin{itemize}
\item Evidence for for $D^0-\overline{D^0}$ mixing has been achieved
  by the Belle experiment \cite{Abe:2007dt}, in the singly Cabibbo
  (SCS) suppressed $D\to K^+K^-,\pi^+\pi^-$ decay modes, and by the
  BaBar experiment \cite{Aubert:2007wf}, in the
  doubly Cabibbo suppressed (DCS) $D\to K\pi$ decay mode.
  \item When combined with previous results from other experiments, 
the signal for $y_{\rm CP}\neq0$ in the SCS decay is strengthened (to
about $4\sigma$). 
\item The evidence implies a width difference at the one percent
  level. In the limit of small CP violation, the CP-odd state is
  longer-lived ($|D_-\rangle=|D_L\rangle,\ |D_+\rangle=|D_S\rangle$).
\item There is no evidence for either mass splitting or CP
  violation.
\item A width difference $y\sim0.01$ is consistent with the Standard
  Model. In particular, it suggests that SU(3) breaking from phase
  space effects, identified and calculated in Ref. \cite{Falk:2001hx},
  are likely to play a major role. In that case, $x$ should be not far
  below present bounds \cite{Falk:2004wg}.
  \item The fact that $|x/y|>1$ seems to be disfavored, and that there
    is not even a hint to CP violation, implies that there is no hint
    for new physics.
    \item The stronger bounds on $x$ and on $A_m$ imply that
      supersymmetric models 
      of alignment are viable only if (i) there is some level of
      squark degeneracy from RGE, and/or (ii) the squark and gluino
      masses are heavier than 2 TeV, and/or (iii) there is
      accidental cancellation between various supersymmetric
      diagrams. The likelihood of observing light non-degenerate
      squarks at the LHC became considerably lower. 
      \end{itemize}

Mixing, CP violation in mixing, and CP violation in the interference of decays
with and without mixing, should affect all neutral $D$-meson decays to
final states that are common to $D^0$ and $\overline{D^0}$. Thus, the
picture that is now emerging -- $y\sim0.01$, $|x|\lsim|y|$ and small or
zero CP violation -- can be further tested and sharpened by additional
experimental results.

While this paper was being written, a related study appeared
\cite{Ciuchini:2007cw}.

\section*{Acknowledgments} 
I am grateful to Bob Cahn and Zoltan Ligeti for useful discussions. I
thank Ikaros Bigi and Zoltan Ligeti for valuable comments on the
manuscript. The research of Y.N.
is supported by grants from the Israel Science Foundation founded by the
Israel Academy of Sciences and Humanities, the United States-Israel
Binational Science Foundation (BSF), Jerusalem, Israel, the
German-Israeli foundation for scientific research and development
(GIF), and the Minerva Foundation.

\appendix
\section{Experimental results}
\label{app:exp}
In addition to the new experimental results that give evidence for
$D^0-\overline{D^0}$ mixing, there is additional data that has not
given such evidence. We here present the relevant data, and combine it
with the new results to obtain world averages.

The experimental results on $y_{\rm CP}$ are the following: 
\beqa\label{ydelybab}
y_{\rm CP}&=&\begin{cases}
  (3.42\pm 1.39 \pm 0.74)\times10^{-2}&{\rm FOCUS\ [35]}\cr
  (0.8 \pm 2.9 \pm 1.0)\times10^{-2}&{\rm E791\  [36]}\cr
  (-1.2 \pm 2.5 \pm 1.4)\times10^{-2}&{\rm CLEO\  [37]}\cr
  (0.8\pm0.4^{+0.5}_{-0.4})\times10^{-2}&{\rm BaBar\ [38]}\cr
  (1.31\pm0.30\pm0.25)\times10^{-2}&{\rm Belle\ [2]}\cr \end{cases}
\eeqa
leading to world average of
\beq\label{ycpwa}
y_{\rm CP}=(+1.20\pm0.31)\times10^{-2}.
\eeq
Thus, the evidence for $y_{\rm CP}\neq0$ is strengthened by other
experiments, and the world average gives it a $4\sigma$ significance.

The experimental results on $\Delta Y(=-A_\Gamma)$ are the following:
\beq\label{deltayexp}
\Delta Y=\begin{cases}
  (-0.8\pm0.6\pm0.2)\times10^{-2}&{\rm BaBar\ [38]}\cr
(-0.01\pm0.30\pm0.15)\times10^{-2}&{\rm Belle\ [2]}\cr \end{cases}
\eeq
leading to world average of
\beq\label{delywa}
\Delta Y=(-0.21\pm0.30)\times10^{-2}.
\eeq

The experimental results on $y^\prime_0=y^\prime\cos\phi$, assuming CP
conservation, are the following:\footnote{We do not include the
  results of E791 \cite{Aitala:1996fg} and FOCUS \cite{Link:2004vk}
  which are not given in a form appropriate for our purposes.}
\beq\label{ypexp}
y^\prime_0=\begin{cases}
  (-23\pm14\pm3)\times10^{-3}&{\rm CLEO\ [41]}\cr
  (0.6^{+4.0}_{-3.9})\times10^{-3}&{\rm Belle\ [12]}\cr
  (9.7\pm4.4\pm3.1)\times10^{-3}&{\rm BaBar\ [1]}\cr
  \end{cases}
\eeq
leading to world average of
\beq\label{ypwa}
y^\prime_0=(2.5\pm3.1)\times10^{-3}.
\eeq
Thus, the data from other experiments (particularly the lower range
measured by Belle) weaken the signal for $y^\prime\neq0$ to below
one sigma. Note, however, that the information on the correlation
between $y^\prime$ and $x^\prime$ is lost in this simple averaging. 

The experimental results on $x^{\prime2}_0$, assuming CP
conservation, are the following:
\beq\label{xpexp}
x^{\prime2}_0=\begin{cases}
  (1.8^{+2.1}_{-2.3})\times10^{-4}&{\rm Belle\ [12]}\cr
  (-2.2\pm3.0\pm2.1)\times10^{-4}&{\rm BaBar\ [1]}\cr
  \end{cases}
\eeq
leading to world average of
\beq\label{xpwa}
x^{\prime2}=(0.7\pm1.9)\times10^{-4}.
\eeq
If we interpret this bound as $|x^\prime|=(0.8\pm0.8)\times10^{-2}$, and
combine that with the CLEO result \cite{Godang:1999yd}
$|x^\prime|=(0.0\pm1.5\pm0.2)\times10^{-2}$, we obtain
\beq
|x^\prime|=(0.6\pm0.7)\times10^{-2}.
\eeq

Finally, we mention significant new (preliminary) results
\cite{staric} on a Dalitz plot analysis that yields
\beqa\label{belledalitz}
x&=&(0.80\pm0.29\pm0.17)\times10^{-2},\no\\
y&=&(0.33\pm0.24\pm0.15)\times10^{-2}.
\eeqa


\end{document}